\documentclass[aip,reprint]{revtex4-1}
\usepackage{graphicx}
\usepackage{svg}
\usepackage{amsmath}
\usepackage{xcolor}
\draft 

\newcommand{\ket}[1]{$|\,#1\rangle$}

\begin{document}

\title{Operational entanglement-based quantum key distribution over 50\,km of real-field optical fibres} 

\author{Yoann Pelet}
\email[]{yoann.pelet@univ-cotedazur.fr}
\affiliation{Universit\'e C\^ote d'Azur, CNRS, Institut de Physique de Nice (INPHYNI), UMR 7010, 06108 Nice Cedex 2, France}
\author{Gr\'egory Sauder}
\affiliation{Universit\'e C\^ote d'Azur, CNRS, Institut de Physique de Nice (INPHYNI), UMR 7010, 06108 Nice Cedex 2, France}
\author{Mathis Cohen}
\affiliation{Universit\'e C\^ote d'Azur, CNRS, Institut de Physique de Nice (INPHYNI), UMR 7010, 06108 Nice Cedex 2, France}
\author{Laurent Labont\'e}
\affiliation{Universit\'e C\^ote d'Azur, CNRS, Institut de Physique de Nice (INPHYNI), UMR 7010, 06108 Nice Cedex 2, France}
\author{Olivier Alibart}
\affiliation{Universit\'e C\^ote d'Azur, CNRS, Institut de Physique de Nice (INPHYNI), UMR 7010, 06108 Nice Cedex 2, France}
\author{Anthony Martin}
\email[]{anthony.martin@univ-cotedazur.fr}
\author{S\'ebastien Tanzilli}
\email[]{sebastien.tanzilli@univ-cotedazur.fr}
\affiliation{Universit\'e C\^ote d'Azur, CNRS, Institut de Physique de Nice (INPHYNI), UMR 7010, 06108 Nice Cedex 2, France}

\date{\today}

\begin{abstract}
We present a real field quantum key distribution link based on energy-time entanglement. Three nodes are connected over the city of Nice by means of optical fibers with a total distance of 50\,km. We have implemented a high-quality source of energy-time entangled photon pairs and actively stabilized analysers to project the quantum states, associated with an innovative remote synchronization method of the end stations' clocks which does not require any dedicated channel. The system is compatible with the ITU 100\,GHz standard telecom-grid, through which a raw key rate of 40\,kbps per pair of channels is obtained. A post-treatment software performs all the necessary post-processing procedures enabling to establish secret keys in real time. All of those embedded systems and achieved performance make this network the first fully operational entanglement based metropolitan quantum network to be implemented in real field.
\end{abstract}


\maketitle 

\section{Introduction}
\noindent
Quantum key distribution (QKD) offers the possibility of sharing unconditionally secure cryptographic keys between multiple distant users. Being one of the most mature quantum application, protocols have improved and diversified to meet all the specific requirements of practical secret key distribution in terms of security, distance, and rate. QKD protocols, purely theoretical at first \cite{bennett1984quantum,bennett_quantum_1992}, were soon tested in laboratory \cite{bennett1992experimental} and have been gradually extended to real field implementations, exploiting  free-space links toward satellites \cite{liao2017satellite} as well as terrestrial and underwater \cite{wengerowsky2019entanglement} fiber links. The improved original QKD protocol (decoy BB84) \cite{hwang2003quantum} is used to communicate through thousands of kilometers with trusted nodes acting as repeaters in China~\cite{chen_integrated_2021}. Twin-field QKD has set the longest distance for a repeater-less link \cite{chen_twin-field_2021}, while entanglement-based protocols (BBM92) can connect many users at a time using various topologies, or reach very high key rate between two users exploiting multiplexing scheme\cite{aktas2016entanglement,appas2021,1GHZ_ursin}. \\

\begin{figure*}[htb]
\begin{center}
\includegraphics[width=1.95\columnwidth]{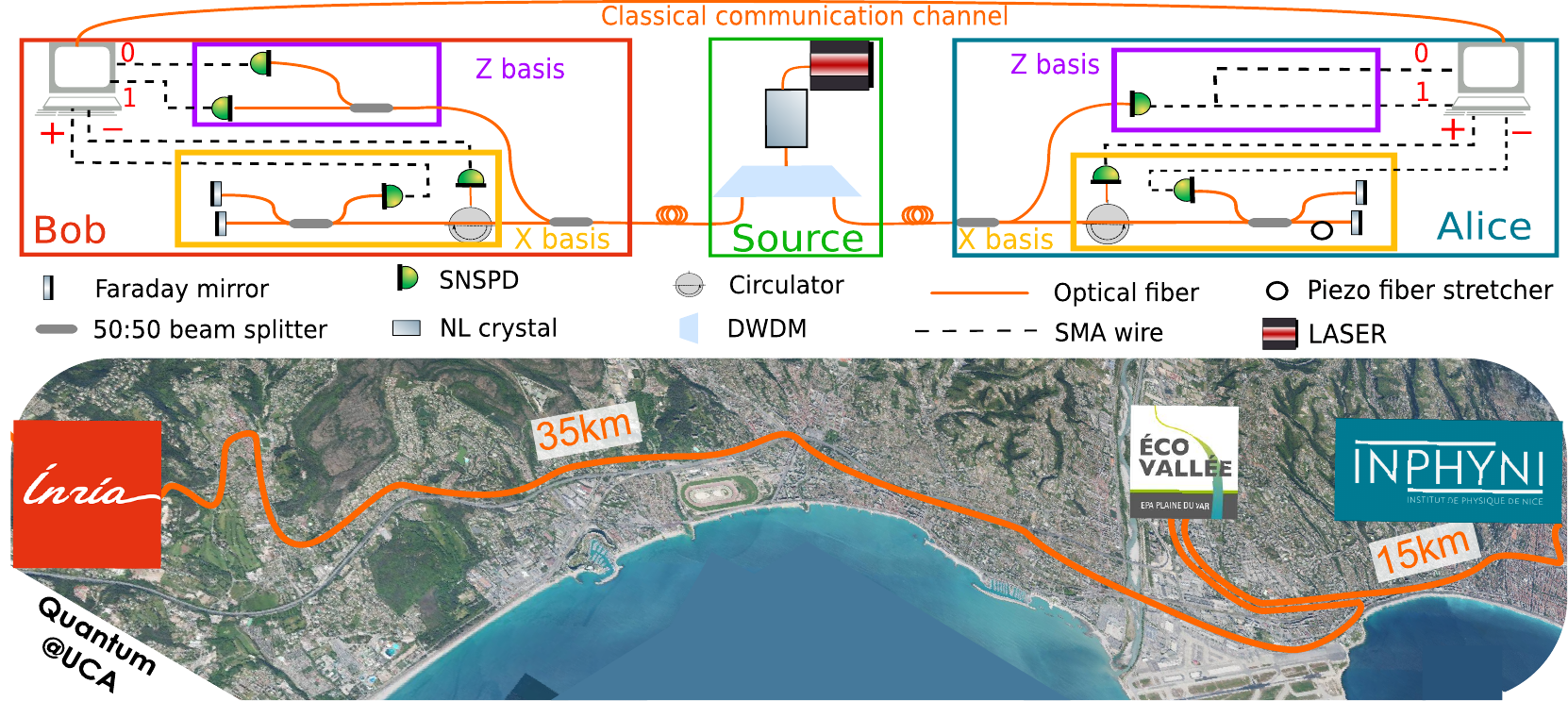}
\caption{Experimental setup. The source, located at the central station, creates energy time bipartite entanglement shared between Alice and Bob (end stations) using DWDM (ITU channels 20 and 22). Each analyser (Alice and Bob) sends randomly the photons in either the Z or the X basis using a $50/50$ beam splitter. The X basis is set up using two, stabilized Michelson interferometers with the same delay, one at Alice's station, the other at Bob's. The Z basis consists of a $50/50$ beam-splitter on Bob's side with a short and a long path to the detectors while Alice only has one detector, short and long paths being created electronically. Everything is fibered, with standard SMF-28 fibers for all components except for the connection between the pump laser and the PPLN waveguide ensured by a polarization maintaining fiber.\label{fig_setup}}
\end{center}
\end{figure*}

As described in Fig \ref{fig_setup}, we present a fully operational QKD link over 50\,km of deployed telecom fiber exploiting energy-time entangled photon pairs distributed to two users (named Alice and Bob) across the M\'etropole C\^ote d'Azur. We measure a raw keyrate of $40\,$kbps with a total of $20.5\,$dB of transmission losses, without limits for the operation time. A fully automated stabilization protocol continuously maintains the quantum bit error rate (QBER) under $7$\%. A complete post treatment program produces a final key rate of $6.5\,$kbps for one pair of ITU channels among the 40 pairs that are available with our multiplexing strategy.\\
There are three original features in our scheme. First, the users' clock synchronization, fundamental in QKD protocols, is carried out using the same qubits transmitted during the QKD protocol, without requiring any additional dedicated channel \cite{ho2009clock}. It allows to actively pace, in real time, the two distant clocks with a precision of $32$\,ps. Second, we exploit a novel, home made, asymmetric sifting protocol increasing by a factor 2 the raw key rate and requiring one less detector compared to similar implementations\cite{neumann2022continuous}. Third, we implement a software performing on-the-fly post-treatment operations required to transform the raw correlated bits into secured secret keys. With those features, our network generates, continuously, actual secret keys ready for practical exploitation of secured communication over more than 32 hours. To our knowledge, this is the first demonstration of a fully geared quantum link up-gradable to a network and the longest uninterrupted run of entanglement-based QKD in real field.

\section{Conceptual operation}
\noindent
\textbf{The photon pairs} are generated via spontaneous parametric down conversion (SPDC) by mean of continuous laser light sent to a non linear crystal. Thanks to the conservation of energy and momentum between the three interacting fields, the emission time of each pair is unknown to the coherence length of the pump laser. The photon pair distribution is poissonnian, guaranteeing true randomness of the emission time.\\
\textbf{The measurements} of the photons' states are performed randomly using two complementary basis, Z and X, with equal probabilities ($1/2$). The basis and time of arrival of the photons are recorded for each detection event. 
The Z basis is used to generate the key and corresponds to the measurement of the arrival time of the photons $\{|\,0\rangle,\,|\,1\rangle\}$.
The X basis is used to estimate the eavesdropper potential information by measuring two-photon interference between the two possible arrival times $\{|\,0\rangle,\,|\,1\rangle\}$, using the coherent superposition $|\pm\rangle = \frac{|\,0\rangle \pm |\,1\rangle}{\sqrt{2}}$.\\
\textbf{Time correlation and basis reconciliation} occurs every 100\,ms.
Bob publicly announces his photons' detection times, measurement basis and, only for the X basis, output bits.
In her data set, Alice searches the time and basis correlated events and sends back to Bob the event occurring in the Z basis. 
Alice and Bob can then filter their detection events to generate blocks of raw key containing $n_{Z}$ events.
Simultaneously, Alice records all coincidence events in the X basis to compute the phases error rates $\phi_x$ contributing to the QBER in the X basis: QBERX.\\
\textbf{The Error correction} protocol is applied over blocks of $n_{Z}$ bits.
During the protocol, the leakage of information (\textit{i.e.}, the number of bits disclosed) is given by $\lambda_{EC} = f_{EC}\,\cdot\, n^{EC}_{Z}\,\cdot\, h(Qz), $ where $f_{EC}$ is the efficiency of reconciliation, $h(x)$ the binary entropy, and $Qz$ the error rate in the Z basis. 
We use a Cascade algorithm developed in Ref.\cite{martinez-mateo_demystifying_2014} with a reconciliation efficiency of 1.06.\\
\textbf{Privacy amplification (PA)} is the last step toward generating a secure key. Here, $k$ error correction blocks are fused to create a PA block.
On each PA block, Alice computes the lower bound over the length $l$ of the secret key taking into account the finite statistic effect, $l< n_Z (1-h(\phi^u_z)) - \lambda_{EC}$, with $n_z = k n^{EC}_Z$ and $\phi^u_z$ the upper bound on the phase error rate estimated with the events in the Z basis.
The extractor is based on a Toeplitz-hashing function extracting $l$ bits from the $n_z$ input bits\cite{bennett_generalized_1995}.

\section{Experimental operation}

\subsection{Source}
\begin{figure}
    \includegraphics[width=9cm]{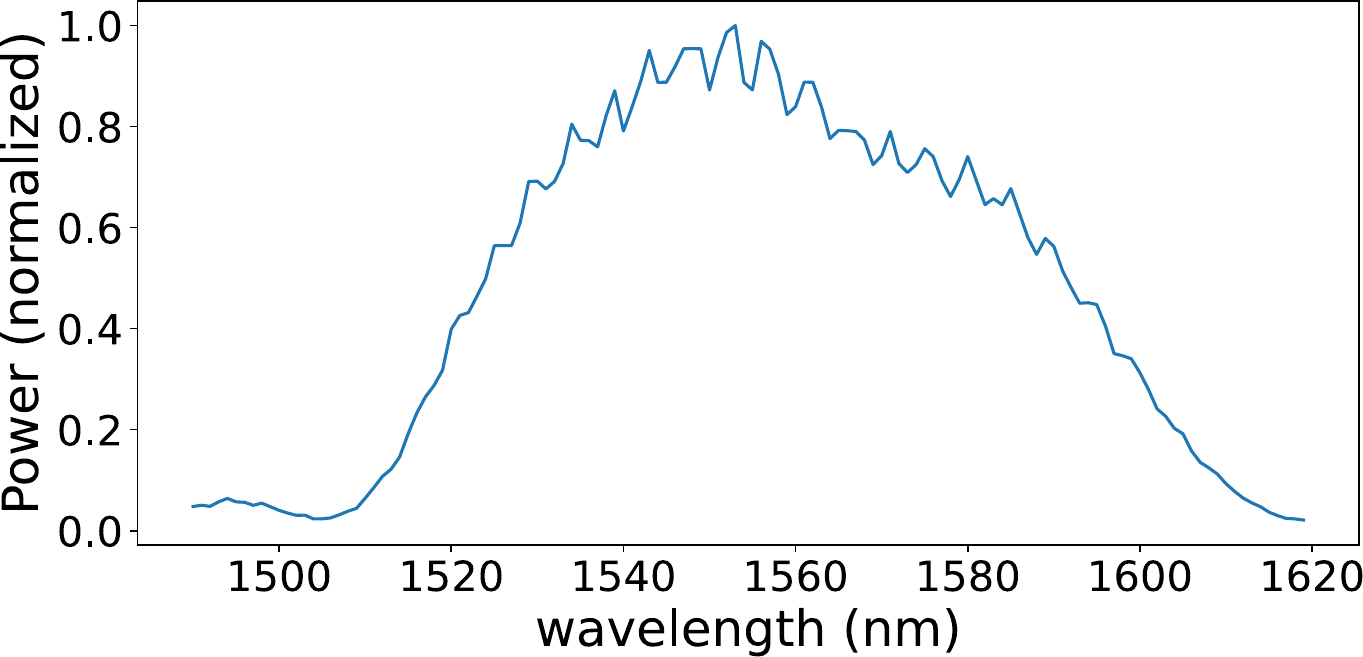}
    \caption{Spectrum of the photon pairs generated from the nonlinear crystal inside of the photon pair source.}
    \label{fig_spectre}
\end{figure}

\noindent
The photon pairs are emitted via spontaneous parametric down-conversion (SPDC) from a continuous-wave (CW) laser at $780,10$\,nm with $1$\,MHz linewidth interacting with a $\chi^{(2)}$ non linear crystal. The latter is a periodically-poled lithium niobate ridge waveguide (PPLN). Thanks to the conservation of the energy and momentum, also called quasi-phase matching in our case, the non-linear interaction permits to produce degenerate paired photons at $1560.20$\,nm over a spectral bandwidth of $80$\,nm as shown in Fig \ref{fig_spectre}. The crystal has a normalized brightness of $1.8*10^{9}$\,pairs/nm/mW with a coupling efficiency of 54.2\% in a standard telecom single-mode fiber.
For the two users demonstration, we exploit one pair of standard 100\,GHz ITU channel dense wavelength demultiplexers (DWDM) around the degeneracy. 
The channel at $\lambda=1560,61\,$nm and at $1559.79\,$nm are sent to Alice and Bob, respectively.

More complex network topologies can be implemented by exploiting the broadband emission of the crystal. 
With the $100\,$GHz ITU channels, 40 independent sources of energy-time entangled photon-pairs can be operated in parallel\cite{aktas2016entanglement}.
This allows either to create a $40$ users fully connected network or to increase the keyrate of a single QKD link by 40 times.
These numbers can go even higher by using $50$\,GHz or $25$\,GHz DWDM.

\subsection{Analysers}
\label{Analysers}
Each analyser allows for a passive basis choice between X and Z (a Michelson interferometer or a straight detection line) thanks to a $50/50$ fibered a beam-splitter, \textit{i.e.} $P_{Z}=P_{X}=\frac{1}{2}$, as shown in \figurename{~\ref{fig_setup}}.

\textbf{The Z basis} decodes the incoming photons using a short and a long path separed by a time delay $\tau$. As shown in Fig \ref{fig_setup} on Bob's side, a $50/50$ beam-splitter connected to detectors "0" and "1", each corresponding to a bit value of the raw key. The delay $\tau$ is added on the arm corresponding to the bit "1". On Alice's side, we employ only one detector connected to two channels of the time to digital converter (TDC) with an electrical delay $\tau$ between the two. Similarly to Bob's side, the channel with a delay corresponds to the bit "1" while the other is reffered to as "0".
Every detection event on Alice's Z basis leads to the measurement of both \ket{0} and \ket{1}. However, comparing both detection times with Bob's timestamps allows Alice to know whether Bob has measured \ket{0} or \ket{1} by post-selecting only the ones with a zero coincidence delay.
To properly distinguish between \ket{0} and \ket{1}, the delay $\tau$ must be larger than the timing jitter of the coincidence detection electronics. 
Usually, time bin encoding is done using two different optical paths of different length leading to the detection of \ket{0} or \ket{1}. 
The choice between these two paths being passive, the probability that both photons take the same path and create an exploitable coincidence is $25$\%. 
Our scheme allows to bypass the passive choice between the two paths on Alice's side, improving to $50$\% the probability of recording a coincidence.

\textbf{The X basis} monitors the visibility of the interference induced by the photon pairs using unbalanced fiber Michelson interferometers.
Two-photon interference happens independently of the incoming polarization by using polarization independent $50/50$ beam splitter together with Faraday mirrors, thus avoiding the need for active polarization stabilization.
For the X basis to be able to guarantee security in the Z basis, the path length difference between the two arms of the interferometer is set to $\tau$, same as the delay used in the Z basis. 
To maximize the two-photon interference visibility, the delay of the interferometers must satisfy the following equations:
\begin{equation}
    \begin{gathered}
        \tau^c_s < \tau_{a,b} < \tau^c_p,\\
        |\tau_{a} - \tau_b| \ll \tau^c_s,
    \end{gathered}
    \label{delay}
\end{equation}
where $\tau_a$ ($\tau_b$) corresponds the delay between the arms of Alice's (Bob's) interferometer, and where $\tau^c_s$ and $\tau^c_p$ are the coherence times of the single photons and the photon pairs, respectively.
In continuous regime, the coherence of the pairs correspond to that of the coherence time of the pump laser ($\approx 300\,$ns) while the coherence of the single photons ($\approx 3\,$fs) is given by the bandwidth of the regarded ITU channel, \textit{i.e.} 100\,GHz.
To satisfy the first relation of eq.\ref{delay}, we chose an interferometer path length difference of 46\,cm, \textit{i.e} $\tau_{a,b} =1.6$\,ns.
The second relation imposes to make the interferometers as identical as possible.
We have $|\tau_{a} - \tau_{b}| < 3\,$fs, \textit{i.e.} a fiber length difference shorter than $100\,{\rm \mu m}$, leading to a raw visibility of $99.7\%$ from the photon pairs.
The two interferometers are set up at remote locations, separated by $50$\,km. Consequently, they need to be temperature stabilized to reduce relative phase fluctuations. 
In addition, a piezo fiber-stretcher is required on the long arm of Alice's interferometer to lock the relative phase between Alice and Bob to an appropriate value.

All photons are detected with superconducting nanowire single photon detectors (SNSPD) IDQuantique ID281 showing a dead-time of $\sim50$\,ns and a detection efficiency of 80\%. 
Timestamps are recorded with two time-to-digital converters (TDC - Swabian Instrument Timetagger Ultra) showing 1\,ps resolution. 
The combination of the detector and TDC jitters give a coincidence peak with a FWHM of 100\,ps. \\ 
The SNSPDs employed in our QKD system have a meander pattern which introduces polarisation-dependent efficiency \cite{zheng2016design}. In fibers, the polarisation rotates over time, which creates instability of the secret key rate generated by the system.
To overcome this issue, two electrical adjustable polarisation controllers are placed at the output of the source and are constantly adjusted to maximize the number of detection events at Alice's and Bob's. 

\subsection{Quantum channel}
\noindent
The QKD protocol is operated over a dedicated optical fiber network deployed over the M\'etropole C\^ote d'Azur.
As shown in Fig \ref{fig_setup}, the analysers at Alice and Bob's are located in Nice and Sophia Antipolis, with the source placed in-between the two sites. 
More specifically, we exploit two standard, telecom, single-mode dark fibers.
The fibers induce $5.7\,$dB and $10.5\,$dB of losses from the source to Alice's or Bob's end station, respectively.
Chromatic dispersion along the fibers broadens the peaks in the coincidence histogram, leading to a decrease of the signal to noise ratio in the coincidence window\cite{fasel2004quantum}.
This makes the use of a  dispersion compensation module mandatory.
Energy-time entangled photon-pair allows nonlocal dispersion cancellation, such that the effect of dispersion on one photon can be canceled out by the dispersion experienced by its correlated photon \cite{steinberg1992dispersion}.
Therefore, a 50\,km fiber dispersion compensation module, inducing $4.3\,$dB of additional losses, is added before Alice's analyser to compensate the dispersion accumulated by the photon-pairs.
By placing the module on Alice's path, the transmission losses between the two users are balanced.

\begin{table}[b]
\begin{tabular}{lcccccccc}
\hline
Detector & AZ & AX1 & AX2 & BZ1 & BZ2 & BX1 & BX2\\
\hline
Losses (dB)& 12.6 & 20.33 & 20.46 & 15.86 &16.22&21.83&22.47\\
\hline
\end{tabular}
\caption{Losses from the source to each detector, with A and B for Alice and Bob and X and Z for the measurement basis.}
\label{losses}
\end{table}

As detailed in table~\ref{losses}, the analysers add on average $11\,$dB before each detector for both X basis, $6\,$dB for Bob's Z basis and $3\,$dB for Alice's Z basis.
The losses between the source and the analysers are standard numbers when it comes to classical telecommunication ($\approx 0.3\,$dB/km). At the analysers, losses are mostly due to the beam splitters, while the interferometers also induce $2\,$dB of extra losses due to the splices and Faraday mirrors. The latter cannot be reduced drastically and are inherent to the protocol itself.
We can assume that any deployed metropolitan quantum network in the future will have to deal with similar characteristics when using standard telecommunication fibers.

\section{Data post-treatment}
\label{sync}
\noindent
All data acquisition and post-treatment operations are automatically performed using a homemade Labview software which performs, in real time, every calculation and stabilization from the detection events to the generation of the secure key on Alice and Bob sides.

Our QKD protocol heavily relies on the precision of the recorded arrival times of the photons at Alice's and Bob's detectors.
To measure those, two independent TDCs are used, one on each site. 
When a detection occurs, the TDC generate a 64 bits time-tag corresponding to the elapsed time between the start of the device and the arrival time of the photon.
The time-tagging precision relies on that of the clock used as a reference.
When using two distant devices, two synchronization problems need to be addressed: i) defining a time zero, and ii) maintaining the same clock rate on both systems.

During the initialization of the QKD system, the laser beam is blocked by an electronically tunable optical attenuator.
When unblocking the light, the SPDC source starts to generate pairs of photons that are subsequently separated and sent simultaneously to Alice and Bob.
On each side, an algorithmic filtering function verifies, every millisecond, if the number of detection is largely greater than the dark-count rate of the detectors.
When the detection rate becomes large enough, the two TDCs send a block of detection acquisitions of 100\,ms. 
In this block, the difference between the two first time-tags recorded by Alice and Bob corresponds to the delay between correlated events. The time tagger functions allow to get this first delay with a precision better than $10\,$ns.
Alice performs a correlation calculation on one set of acquisition blocks around the delay found with the first time-tag event and looks for the coincidence peak.
The identification of this peak allows to define the zero delay between the two TDCs with a precision of about $1$\,ns. 
Once this is done, the time window for the correlation can be reduced to $8$0\,ns and the precision of the zero delay measurement improves to $32$\,ps for the next blocks of raw keys.

While absolutely necessary, getting a precise initialization time does not guarantee an everlasting synchronization.
Two independent clocks drift apart over time. However, our system requires at all times, a clock synchronization better than the coincidence window used to validate correlated detection events which is set at $128$\,ps.
The Time-Taggers internal clocks show a drift of $100\,$ns/s. When replaced by a rubidium atomic clocks (Spectratime LNRClock 1500), the drift lowers to $100\,$ps/s.
Even though, the cumulative drift over one day is largely greater than the coincidence window. 24/7 operation of the link requires active synchronization of Alice's and Bob's clocks.
To do so, a new coincidence calculation is performed every second, to measure the evolution of the zero delay and therefore to adjust the internal frequency of the Alice's clock.
This active correction ensures that the clocks' synchronization is always better than $32$\,ps.

This kind of stabilization does not require any dedicated channel\cite{chen_integrated_2021}. The drift measurement is straightforwardly done by exploiting the entanglement resource used to perform the QKD protocol and does not lower the keyrate.
It also allows to comply with the optical length fluctuation of the quantum channel.
All those perturbations normally affect the synchronization over long time scales but are here simply and constantly corrected.

For each acquisition block received from Bob, Alice performs the sifting to extract the time tags and basis of all the correlated events. Alice sends back to Bob the relevant events obtained in the Z basis to allow the generation of the key, while she keeps the events in the X basis to compute the QBERx .
The events in the Z basis are accumulated to generate a correction block of $n^c_{Z} = 16384$ bits and send to the error correction function based on a cascade algorithm.
The QBERx is calculated for each correction block and the value is used as input of a feedback loop algorithm minimizing the QBERx by changing the voltage applied to the piezo fiber-stretcher located in Alice's interferometer.
Lastly, after the accumulation of 100 correction blocks, \textit{i.e.} $n_z = 1.6384 \times 10^6$, the secret key length is computed taking into account the finite statistic effect via the security proof described in Ref \cite{cai2009finite}. 
The corrected key is then sent to a privacy amplification function based on a Toeplitz-hashing extractor to compute the final secret key on both sides.  
All operations from data acquisition to privacy amplification are performed simultaneously between Alice's and Bob's computers, allowing the link to operate continuously without any down-time required for post-treatment.

\section{Results}

\subsection{Optimization of the secret key rate}
\label{Simulation}
\begin{figure}
    \includegraphics[width=9cm]{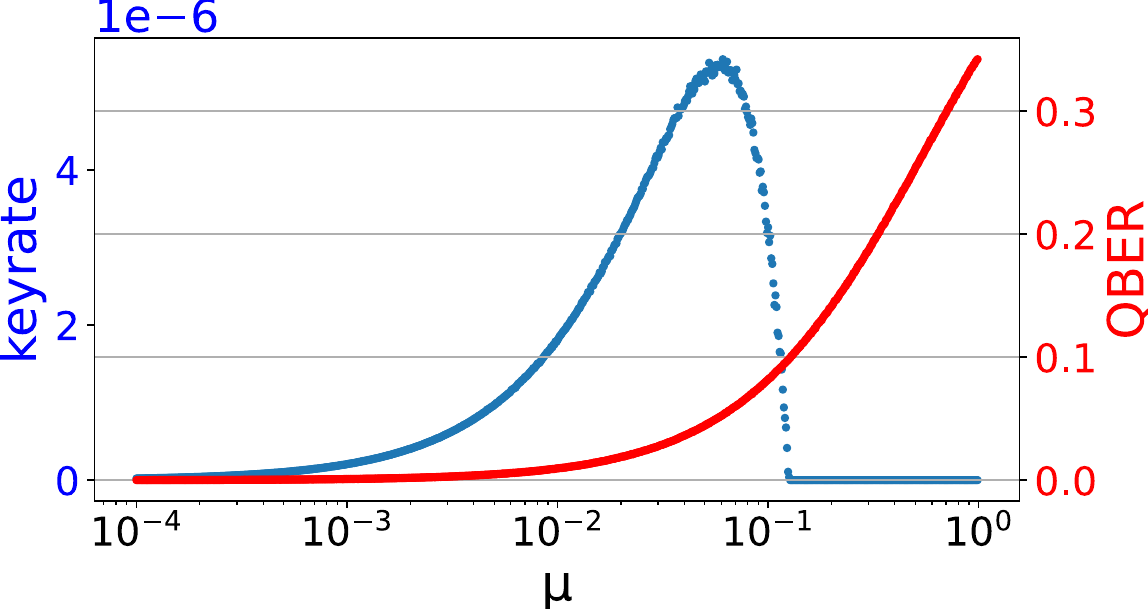}
    \caption{Simulated SKR (blue) and QBERz (red) as a function of $\mu$ : the number of photon pairs outputing from the source per time wondow ($128\,$ps).}
    \label{simu}
\end{figure}

\noindent
The value of the SKR (Secret Key Rate) results from a competition between the number of entangled photons detected and the error rate in the recorded detection events. 
These values depend on several parameters: the photon-pairs generation probability, the losses, the detectors' dark counts, the size of the coincidence windows and the imperfection in the source and the analysers. 
After characterising the losses induced by the transmission of the photons from the source to both users, those induced by the analysers and the intrinsic errors of the system, we have simulated the evolution of the SKR and QBERz as a function of the photon-pair generation probability per windows of interest, as shown in \figurename{\,\ref{simu}}.
We choose a coincidence window in the post-treatment of $128\,$ps, which is the smallest size we can set before loosing too much coincidence events due to detection timing-jitter and residual clock drift. 

Increasing the pump power augments the amount of information shared per second, resulting in a higher SKR. 
However, the higher the power, the higher the probability of generating double pairs of photons (two different pairs emitted within the same time window), which increases the QBER.
To ensure unconditional security, all errors in the detection events have to be considered as potential eavesdropper's actions.
Therefore, to compensate for the double pair induced errors, more bits of the raw key have to be sacrificed to create the secret key resulting in a decrease of the SKR. 
Therefore, the optimal SKR is found for the highest pump power before reaching a critical rate of double pairs. 
We find the optimal value for the QBER to be $5.0$\% in the Z basis for an optimal SKR.

\begin{figure}
    \includegraphics[width=9cm]{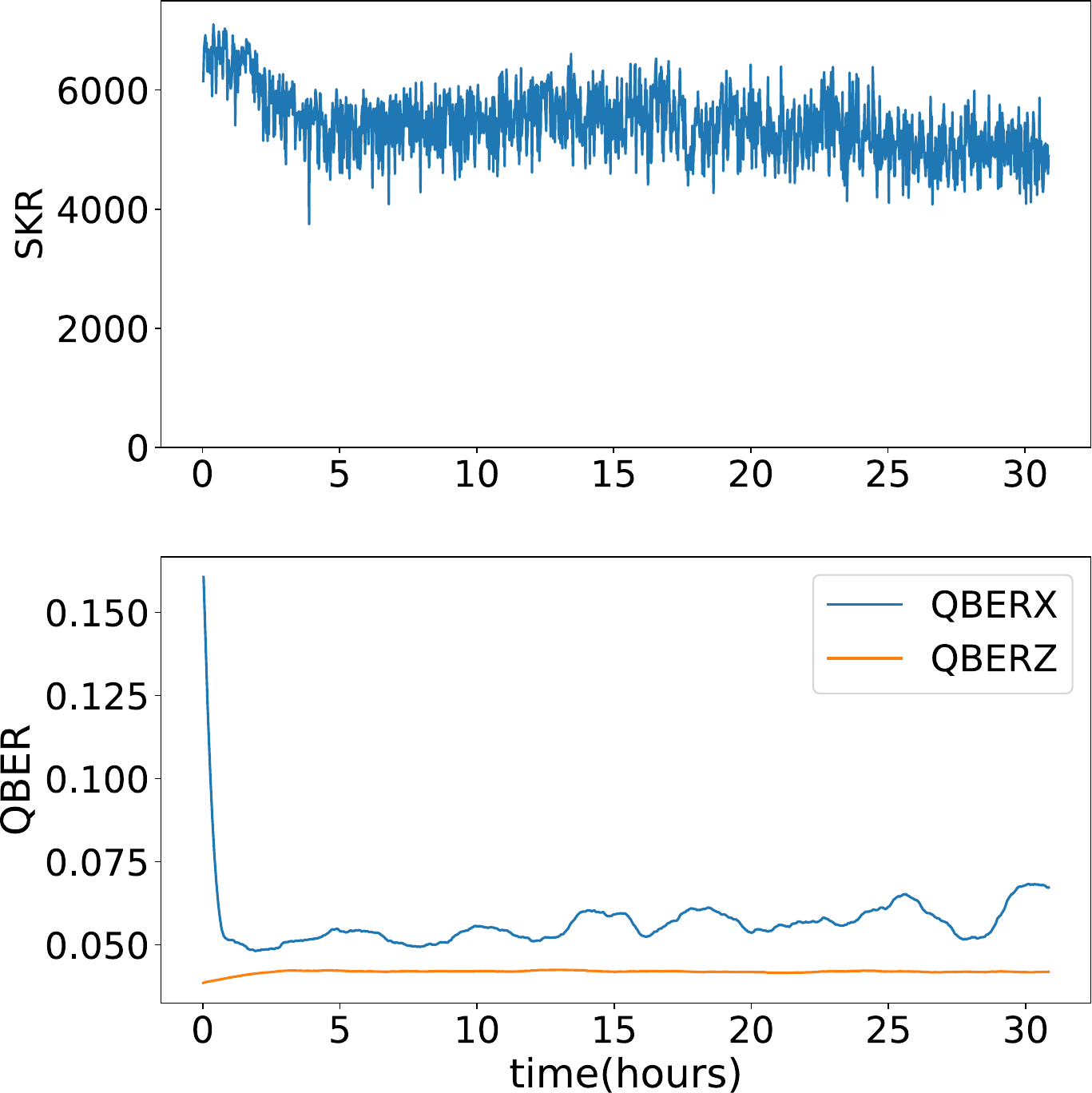}
    \caption{Secret Key Rate (top) and QBER in both basis (bottom) as a function of time for a 30h uninterrupted run.}
    \label{keyrate}
\end{figure}

As shown in \figurename{\,\ref{keyrate}}, the results obtained experimentally in the optimal configuration lead to an average SKR of $6.5\,$kbps. 
We perform a continuous measurement lasting $31\,$h, before a helium condensation cycle is necessary for our SNSPD system.
As we can see, all feedback loops of the system act on the different devices to maintain both the QBER and SKR during the entire operation time.
The SKR depicted here corresponds to an experimental value of secrets bits stored on a hard drive on each user's computer than can be directly used as cryptographic keys.
To our knowledge, this makes our implementation the first fully automated and operational deployed entanglement-based quantum link. 

\section{Conclusion}
\noindent
We have demonstrated a fully functional and automated quantum key distribution system, operated over 50\,km of deployed standard telecom fibers, allowing to establish shared secret keys between two distant users.
All processing steps required to perform the key distillation protocol are implemented on a single computer on each site. This is, to our knowledge, the first implementation of an entanglement-based QKD protocol with automated and continuous post-treatment and synchronization. To reach such a level of operationability, we have also achieved a simple and efficient way to synchronize distant users without requiring extra physical channel or specific dedicated data. The secure keyrate reached over the link is $6.5\,$kbps on average for a single pair of DWDM channels and could theoretically reach $7$\,kbps with ideal stabilization processes. Lastly, more users can be added to the link by connecting them to the source using multiplexing strategies, leading to more complex and interestong topologies. This physical configuration would allow to make a fully connected network with up to 40 users using only standard $100$\,GHz DWDM\cite{koduru_trusted_2020}. 


\section*{Data availability}
\noindent
Data are available from the authors on reasonable request.

\section*{Acknowledgements}
\noindent
This work has been conducted within the framework of the French government financial support managed by the Agence Nationale de la Recherche (ANR), within its Investments for the Future programme, under the Universit\'e C\^ote d'Azur UCA-JEDI project (Quantum@UCA, ANR-15-IDEX-01), and under the Strat\'egie Nationale Quantique through the PEPR QComtestbeds project (ANR 22-PETQ-0011). This work has also been conducted within the framework of the OPTIMAL project, funded by the European Union and the Conseil R\'egional SUD-PACA by means of the 'Fonds Europ\'eens de d\'eveloppement regional' (FEDER). The authors also acknowledge financial support from the Conseil R\'egional SUD-PACA through the INTRIQUE (APEX2019) and SIPS (Apex 2021) projects. The authors are grateful to S. Canard, A. Ouorou, L. Chotard, L. Londeix from Orange \& Orange Labs for their support, and to Orange for the installation and the connection of the dark fibers between the three different sites of our network, as well as for all the support they provided for their characterization. The authors also thank the M\'etropole Nice C\^ote d'Azur and the Inria Centre at Universit\'e C\^ote d'Azur for the access to their buildings and for their continuous help in making this network a reality. The authors also acknowledge IDQuantique and Swabian Instruments GmbH teams for all the technical support and the development of new features that were needed for the implementation of our operational QKD system and related experiment. Finally, Y. Pelet acknowledges PhD funding from Accenture and Universit\'e C\^ote d'Azur.

\section*{Author Contributions}
\noindent
Y.P., G.S., L.L., and A.M. performed the experiments. A.M., O.A., L.L., and S.T. designed the protocols and related experiments. G.S. and A.M. performed all software and hardware implementation. Y.P., M.C., and A.M. performed the numerical simulations. Y.P., LL., O.A., A.M., and S.T. wrote the paper. A.M., O.A., and S.T. supervised the work.

\section*{Competing Interests}
\noindent
The authors declare no competing interests.

\end{document}